# Title Page

Title: Bridging the Regulatory Divide: Ensuring Safety and Equity in Wearable Health Technologies

Authors: Akshay Kelshiker (1), Susan Cheng‡ (2), Jivan Achar‡ (3), Leo Anthony Celi* (4,5,6), Divya Jain‡ (7,8), Thinh Nguyen‡ (9), Harsh Patel‡ (10), Nina Prakash‡ (11), Alice Wong‡ (11), Barbara Evans (9)

Statement of Contributions:

‡ All the co-authors contributed equally to the best of their ability.

Corresponding Author: Leo A. Celi. Email: lceli@mit.edu.

Affiliations:

1. Albert Einstein College of Medicine
2. Independent Researcher
3. Case Western Reserve University School of Medicine
4. Massachusetts Institute of Technology
5. Beth Israel Deaconess Medical Center
6. Harvard T.H. Chan School of Public Health
7. University of Washington School of Pharmacy
8. Genentech
9. University of Florida Levin College of Law
10. Medical University of South Carolina
11. Dartmouth College

**Abstract (113 words):**

As wearable health technologies have grown more sophisticated, the distinction between "wellness" and "medical" devices has become increasingly blurred. While some features undergo formal U.S. Food and Drug Administration (FDA) review, many over-the-counter tools operate in a regulatory grey zone, leveraging health-related data and outputs without clinical validation. Further complicating the issue is the widespread repurposing of wellness devices for medical uses, which can introduce safety risks beyond the reach of current oversight. Drawing on legal analysis, case studies, and ethical considerations, we propose an approach emphasizing distributed risk, patient-centered outcomes, and iterative reform. Without a more pluralistic and evolving framework, the promise of wearable health technology risks being undermined by growing inequities, misuse, and eroded public trust.

**Introduction:**

On July 14, 2025, the FDA issued a warning letter to WHOOP, Inc., asserting that its "Blood Pressure Insights" feature, originally marketed as a general wellness tool, actually functions as a medical device and thus violates current FDA regulations [1,2]. The device, which claims to offer "a new way to understand how blood pressure affects performance and well-being," allowed it to bypass regulatory oversight under the guise of "wellness." This enforcement underscores a central challenge in the regulation of wearable health technologies: the growing tension between how wearable products are marketed and how they are actually used.

Wearable health technologies, from fitness trackers to AI-enabled applications, are transforming personal health management. However, regulatory frameworks have struggled to keep pace. Under current law, FDA oversight is determined by the sponsor's "objective intent"—how a product is labeled and marketed—rather than how it is ultimately used [3]. As a result, products like WHOOP's that are labeled as wellness, even if used in high-risk or quasi-clinical contexts, can escape oversight simply by avoiding medical claims. This regulatory gap is further complicated by consumers repurposing wellness devices for medical uses, a practice that remains largely outside the FDA's regulatory reach and can introduce significant safety risks. Such uses can introduce significant safety risk, especially when consumer-grade tools begin informing decisions traditionally guided by medical professionals.

This paper examines the evolving landscape of wearable health technologies, highlighting the regulatory ambiguities and risks created by rapid innovation and shifting consumer use. We analyze the limitations of current oversight mechanisms and propose next steps to ensure that as these tools grow more powerful and widely adopted, they are held to standards of safety, accuracy, privacy, and accountability.

---

**Background on Wearable Devices:**

### a. Evolution

The evolution of wearable health technologies has experienced a recent boom in developments with integrative technologies of the 2010s. Largely fueled by the private sector, wearable health technologies have transitioned from basic fitness trackers to sophisticated medical devices that play a key role in health monitoring. As more technologies have been developed over the decades, the scope of each device has increased drastically.

The late 2000s saw the rise of fitness trackers, such as the Fitbit Classic. This device popularized step counting, heart rate monitoring, and sleep analysis. Individuals were empowered to proactively manage their health through small, readily-available devices [4]. In the mid-2010s, Google Glass pioneered smart glasses that were meant to serve as an accompaniment to other smart devices, like the smartphone or laptop [4]. More recently, smart contact lenses are being developed that can monitor physiological parameters such as glucose and lactate through non-invasive means [5].

As wearable technologies have grown more sophisticated and clinically capable, questions of oversight have become increasingly salient. Understanding how the regulatory system classifies and governs these devices is essential to assessing whether current frameworks are adequate for today's innovations.

### b. Existing Regulatory Classifications of Medical Devices

The Federal Food, Drug, and Cosmetic (FD&C) Act classifies medical devices intended for human use into three categories: Class I, II, and III.

**Class I**
Class I medical devices (e.g. bandages, nonelectric wheelchairs) are those for which general controls alone are sufficient to ensure the safety and effectiveness of the device [6,7]. General controls apply to all device classes but are the sole requirement for Class I. These include good manufacturing practices, accurate labeling, and adverse event reporting. Class I devices must pose minimal risk and are not intended to sustain human life [6].

**Class II**
Class II medical devices (e.g. computerized tomography scanner, infusion pumps for intravenous medications) require both general and special control to ensure safety and effectiveness [7,8]. Special controls may involve postmarket surveillance, patient registries, or dissemination of guidelines. Most Class II devices undergo Premarket Notification, known as the 510(k) process, in which the FDA determines whether the device is "substantially equivalent" to a legally marketed predicate device [9]. If there is enough of an experience base to develop meaningful controls, the device is classified as Class II. Devices without a predicate may be classified as Class I or II through the De Novo pathway, which provides a route for novel, moderate-risk devices without requiring full Premarket Approval [10].

**Class III**
Class III medical devices (e.g. pacemakers, deep-brain stimulators) are those that hold substantial importance in preventing impairment of human health or present an unreasonable risk of illness or injury [7,11]. These devices are subject to Premarket Approval (PMA), a process that includes, but is not limited to, full reports of information to show device safety, a description of device components, and samples of the device for review by the Secretary of Health of Human Services [12]. Unlike the 510(k) pathway, which focuses on equivalence to an existing product, PMA demands substantial demonstration of safety and efficacy.

    c.  **Wellness Devices and Regulatory Ambiguity**

The FDA's Center for Devices and Radiological Health (CDRH) defines general wellness products as those that are solely intended for general wellness use and pose a low risk to the safety of users and others [13]. These products may make claims related to weight management, sleep, or physical fitness. However, they may not make claims to treat or diagnose specific diseases or conditions (e.g. treating obesity). As a result, low-risk wellness products are not regulated under the FD&C Act and are exempt from FDA oversight [13].

In contrast, medical devices that claim to diagnose, treat, or prevent disease are subject to FDA oversight and are classified into Class I, II, or III based on risk level and regulatory burden. This classification determines whether a device must undergo 510(k) clearance, De Novo review, or PMA. This distinction allows companies to bypass clinical validation and reduce time to market by carefully framing their product's objective intent [3,14]. Manufacturers may be incentivized to emphasize a device's benefit to bolster sales while avoiding explicit disease-related claims, allowing even sophisticated AI-driven tools to evade FDA regulation if marketed as wellness products [15].

This creates a possible regulatory blind spot, allowing consumer-facing tools to operate in high-risk contexts without undergoing formal validation or surveillance. Unlike medical devices, which are subject to mandatory adverse reporting and recall systems, wellness devices typically rely on consumer complaints or voluntary recalls [14]. Consequently, errors may go unnoticed until widespread consumer complaints arise.

Further complicating regulation is the issue of third-party repurposing. While the FDA can oversee manufacturers, it has no authority over how consumers use or modify devices post-sale [15]. As *Paek et al.* discuss, users may modify or extend wellness devices using machine learning (ML) algorithms, open-source software, or custom interfaces to derive medical-grade diagnoses, treatments, and insights. In one case in 2016, patients unable to afford hearing aids repurposed consumer-grade personal sound amplification products to meet their medical needs.

These practices illustrate the limitations of current frameworks: even perfect compliance by manufacturers cannot prevent real-world misuse.

### d. Real-world Harms

Such regulatory gaps have tangible, real-world consequences. Emerging research demonstrates that both wearable wellness products and FDA-cleared medical devices can pose risks, challenging the assumption that current classification systems sufficiently protect users and public health.

In 2022, Fitbit recalled the Ionic smartwatch over incidents related to the lithium-ion battery overheating, causing third- and second-degree burns [16]. Because the Ionic wearable was marketed as a wellness device, it bypassed many of the safety checks required of medical devices, including more stringent safety testing, battery performance standards, and monitoring required of FDA-regulated devices. The subsequent recall, after over 100 reports, illustrates a critical flaw in the existing reactionary framework, where wearable devices capable of causing bodily harm can escape essential safeguards by simply avoiding medical claims. Additionally, a recent study demonstrated that many consumer-grade wearable watch bands contain high concentrations of per- and polyfluroalkyl substances (PFAS), specifically perflurohexanoic acid (PFHxA) [17]. These chemicals are applied during the production process to impart water and oil resistance, but the high amounts worn directly on the skin risks dermal exposure and long-term health effects on the liver, hematopoiesis, and the endocrine system.

Wearable technology has also been found to play a role in non-physical harm. Wearable wellness and FDA-regulated devices have been associated with increased preoccupation with symptoms, elevated health anxiety, and greater reliance on informal care pathways among cardiology patients [18]. Furthermore, growing evidence highlights disparities in device accuracy across skin tones, with certain wearable sensors performing less reliably on individuals with darker pigmentation, which exacerbate health disparities and misinform clinical decisions [19]. Psychological distress and biased outputs carry real-world consequences, especially when unregulated devices are used to guide health behaviors, as they often are [20].

Additionally, wearable technologies can also influence public health. A recent study raises concerns about the potential of smart scales, watches, and rings to interfere with cardiac implantable electronic devices [21]. Specifically, certain consumer-grade wearables employ bioimpedance sensing, a noninvasive means of assessing body composition, and these electrical currents are capable of disrupting the normal functioning of pacemakers and implantable cardioverter-defibrillators [22]. These tangible harms underscore the broader ethical and social implications of wearable health technology regulation, which we explore next.

## Ethical and Social Implications/Problems:

**Bias and Equity**

In addition to the consumer safety concerns highlighted above, potential bias in wearables, especially AI-enabled wearables, presents a significant health equity challenge [23,24]. These devices often rely on algorithms trained on data sets that may not accurately represent diverse populations, leading to disparities in accuracy and effectiveness. This inherent bias can exacerbate existing health disparities by providing less reliable information to marginalized groups. These concerns persist even as AI advances rapidly and Large Language Models (LLMs) start to achieve human-level (or better) performance on some metrics. Additionally, LLMs can exhibit "inflexible reasoning," particularly in novel or atypical scenarios [25]. This means that when faced with health data that deviates from their training data, LLMs may struggle to adapt and provide accurate interpretations. The tendency of LLMs to rely on pattern matching from their training data, rather than engaging in flexible, context-aware reasoning, poses a significant risk in healthcare applications, where individual variations and unique circumstances are prevalent.

Notably, the ambiguity of wellness device regulation can present a particular challenge to this area. Wellness device users are a self-selected population, which can skew towards younger, more educated, wealthier individuals of higher socioeconomic status [26]. This demographic bias in user adoption can lead to a skewed understanding of health trends and needs, potentially neglecting the health concerns of underrepresented populations. The fine line between wellness and medical devices further complicates this issue, as data collected during the "wellness" phase may be used to inform later medical applications [27]. This can perpetuate existing biases, as the initial data set may not be representative of the broader population, leading to inequitable outcomes in subsequent medical interventions.

**Data Privacy**

Additionally, data privacy is a paramount concern with wellness devices. While medical devices are subject to stringent regulations to protect patient data, wellness devices often operate outside these frameworks. This can leave users vulnerable to data breaches, unauthorized data sharing, and misuse of personal health information. The average user may be unaware of these distinctions and the associated privacy risks, assuming that their data is protected by the same safeguards as traditional medical devices. The collection of sensitive health data, including heart rate, sleep patterns, and activity levels, without robust privacy protections can have undesired consequences, including discrimination based on health status and the potential for targeted advertising based on personal vulnerabilities [26].

**Profit Motives and Patient Needs**

Finally, the primary focus of data use in the wellness sector may be driven by profit maximization rather than clinical benefit [26,27]. This can lead to the development of products and services that prioritize commercial interests over patient well-being, potentially compromising the quality and effectiveness of health-related interventions. For example, the increasingly high cost of training AI models can create a financial disincentive for companies to continue to refine their models, even after problems with the model have surfaced [28]. The potential for data to be used for targeted advertising or other non-clinical purposes further underscores the need for greater regulatory oversight to ensure that data is used ethically and responsibly.

**Structural Constraints on State-Specific Reform**:

Regulatory frameworks must operate within the bounds of constitutional law, including the limitations on state authority. One key doctrine, the Dormant Commerce Clause (DCC), restricts states from enacting laws that unduly restrict interstate commerce, even in the absence of congressional legislation [29,30]. Given that wearable devices are distributed nationally, state-level regulation of these technologies could trigger DCC challenges. Additionally, state-led regulation introduces concerns; state agencies may lack the technical capacity to regulate wearable devices, especially those involving AI, continuous physiological monitoring, or cloud-based infrastructures. There is also a risk of regulatory capture, where state policymakers are swayed by local interest groups or political considerations, potentially leading to enforcement that disadvantages out-of-state companies.

Currently, states have a few means available to respond to product harm. One is product liability laws, allowing consumers to sue for damages resulting from defective devices [31]. Product liability laws are not federally regulated and are therefore not uniform. As a result, the jurisdiction will affect the types of product liability claims. Additionally, consumer privacy laws provide avenues for remediation in the event of data breaches or unauthorized disclosures; however, these laws are similarly not uniform, as there is no federal law in place [32]. Both these mechanisms, however, are fundamentally reactive, employed only after harm has occurred. Relying solely on post-hoc remedies fails to address the systemic and anticipatory oversight that these technologies require.

**Proposed Solution: A Community-Centered Evolving Regulatory Framework**

While these structural constraints present difficulties to finding solutions, they also highlight the need for an approach that functions within existing boundaries yet remains dynamic, iterative, and participatory — capable of evolving alongside technological advancements and their real-world implications.

A key challenge is the increasing ambiguity between "wellness" and "medical" uses of technology. Existing regulatory paradigms, particularly the FDA's reliance on sponsor "objective intent" to distinguish regulated medical devices from unregulated wellness products, are no longer sufficient in an environment where consumer-facing tools routinely generate health-related insights with clinical significance [33]. Moreover, mapping regulatory frameworks reveals not a binary but a multidimensional continuum, where data may be framed as wellness-related yet carry substantial medical risks, or conversely, where ostensibly health-focused outputs evade formal oversight.

Additionally, no regulatory approach will be complete without holistically addressing the repurposing issue [15]. A more refined strategy will grapple with the reality that consumer interpretation often matters as much, if not more, than marketing claims in determining the true risk profile of these technologies. While the FDA's legal tools are limited in managing consumer use, there are possible future regulatory strategies that can be targeted at end users. For instance, clear consumer warnings, opt-in registration systems that provide educational resources on appropriate device use, or even insurance-linked incentives that reward verified use of clinically-validated features. These downstream measures, in addition to greater oversight over manufacturers and their processes, will help ensure real-world safety for wearable devices.

Furthermore, effective regulation must attend to both the input and output dimensions of AI-driven health technologies. Input regulation concerns the integrity and privacy of the data used to develop and power these tools. It requires ensuring that data collection, storage, and model training practices uphold robust standards of de-identification, privacy protection, and compliance with laws such as HIPAA. However, input safeguards alone are insufficient. Equally critical is output regulation: oversight of how AI-generated information is conveyed to users, including how outputs are framed, the degree to which they are presented as diagnostic, predictive, or therapeutic, and the potential for such outputs to influence high-risk health behaviors.

In navigating these complexities, risk assessment must expand beyond agency-driven determinations. Historically, regulatory risk evaluations have been conducted by centralized bodies based on defined statutory standards. Yet given the diffuse and context-sensitive nature of emerging health technologies, we propose that risk determination must increasingly be community-sourced and community-operated. Diverse patient populations, healthcare professionals, data scientists, ethicists, and advocacy groups must be integrally involved in assessing the harms, benefits, and equity implications of wearable health tools. Importantly, this distributed approach to risk evaluation acknowledges the limitations of any single institution's vantage point and seeks to incorporate the lived experiences and perspectives of those most directly affected by regulatory decisions.

Moreover, patient outcomes must become a primary benchmark for regulatory success. If regulatory frameworks, however well-intentioned, lead to deteriorations in patient health, increased disparities, or reduced trust in medical technologies, such frameworks must be revisited and reformed. Regulatory science must therefore embrace an iterative ethos, grounded in empirical evidence and capable of self-correction.

Another vital pillar of a new regulatory paradigm is the recognition that AI systems must operate within human-centered healthcare models. The promise of AI in augmenting diagnostic and therapeutic decision-making cannot obscure the risks associated with automation bias, inattentional blindness, and overreliance on algorithmic outputs. Thus, AI should be designed not as a replacement for human clinicians but as a tool integrated within regulated medical practice, with clear lines of accountability and robust systems for error detection and correction. This shift will also require investment in clinician training, equipping healthcare providers with the skills to critically appraise and responsibly deploy AI technologies in practice.

Ultimately, we contend that the future of regulating wearable health technologies lies in embracing a distributed and evolving regulatory ecosystem. Universities, health systems, patient advocacy organizations, professional societies, and regulatory bodies must collaborate in ongoing processes of evaluation, standard-setting, and reform. In a social context often characterized by individualism and institutional fragmentation, fostering such collaborative models will be challenging. Nevertheless, it is essential for safeguarding public health, ensuring technological accountability, and promoting equitable access to the benefits of innovation. The task ahead is not simply to construct a new regulatory regime, but to cultivate a regulatory culture that is inclusive, adaptive, evidence-driven, and attuned to the complex realities of 21st-century health technologies.

---

**Future Directions:**

To contend with the structural limitations whilst pursuing improvements to wearable device safety, future efforts should prioritize three key areas: AI-driven personalization, interoperability with medical systems, and policy resilience amid shifting political landscapes.

**AI-Driven Personalization and Predictive Accuracy**

Advances in AI and ML have the potential to significantly improve the predictive accuracy and personalization of wellness technologies. Future research should focus on refining AI models to enhance early detection capabilities, reduce biases in health predictions, and provide more reliable health insights. Establishment of new validation processes may be necessary to accommodate AI's continuous learning capabilities, ensuring that AI-integrated wearables remain safe and effective over time.

The FDA recently published a draft guidance titled, "Considerations for the Use of Artificial Intelligence to Support Regulatory Decision Making for Drug and Biological Products," which provides recommendations to industry on the use of AI to generate data intended to support regulatory decision-making regarding safety, effectiveness, or quality for drugs [34]. Similar regulations and guidance will aid industries in efficiently and safely using AI in their respective fields.

However, in the near term, developers of general-purpose software like LLMs or wellness tools are unlikely to fall under FDA authority unless they overtly claim a clinical use. In future regulatory approaches, stakeholders should consider requiring manufacturers to include clear disclaimers (e.g., "not intended to diagnose, treat, or prevent disease") and resist efforts to impose broad FDA mandates without first demonstrating evidence of harm, as required by the 21st Century Cures Act.

**Data Integration with Healthcare Systems**

Many wellness products "skate the line" between being labeled as devices and thereby dealing with regulatory burdens [27]. Wellness products are generally cheaper to bring to market as well, as they do not have to deal with quality assurance checks. This is not necessarily completely the fault of the manufacturers, but with the flexibility comes less protection and less of a guarantee. Enabling data exchange between wellness devices, regulated medical devices, and electronic health records will be helpful for recommendations, rather than diagnoses, related to health. While technology may one day improve such that AI has the ability to diagnose and prescribe, current work should focus on lighter suggestions – personalized fitness plans rather than insulin dosages, for example. Standardized data-sharing protocols could help bridge the gap between consumer-grade and clinical technologies, allowing healthcare providers to integrate wearable-generated insights into patient care. Future work should explore the development of secure, interoperable platforms that balance accessibility with stringent data privacy protections, especially given the regulatory disparities between wellness and medical technologies.

**Regulatory Adaptation in a Changing Political Landscape**

Regulation is sometimes viewed as a one-time decision, but as the world changes, past policies can become mismatches with emerging new political landscapes [35]. With the potential for significant policy shifts, regulatory pathways for wearable health technologies may face new challenges. Policymakers and industry leaders should proactively explore strategies to safeguard consumer protections while streamlining approval processes for emerging technologies. Developing bipartisan, industry-backed regulatory frameworks could help mitigate potential policy reversals and ensure long-term stability in the oversight of AI-driven wellness devices.

As wearable technologies continue to blur the line between wellness and medical applications, a proactive and adaptive regulatory approach will be essential. Future research should aim to refine AI validation protocols, enhance data interoperability, and anticipate shifts in federal oversight, ultimately fostering an ecosystem that balances innovation, safety, and accessibility.

**Conclusion**:

Wearable health technologies offer the potential to support personalized, data-driven healthcare. However, as these tools grow increasingly sophisticated, their oversight has failed to keep pace. The distinction between wellness and medical devices has become increasingly blurry, and the FDA's reliance on "objective intent" creates loopholes that allow tools to evade regulation, while the patchwork nature of state remedies limits preemptive action. To close this regulatory gap, we argue for a pluralistic, community-informed framework that integrates real-world evidence, patient outcomes, and distributed expertise. This new model must embrace upstream and downstream regulation, address the repurposing and equity challenges, and evolve alongside technologies. As regulation becomes an ongoing interactive process, wearable health technologies will realize their promise without compromising safety, privacy, or equity.